# Institutionalization of Software Product Line: An Empirical Investigation of Key Organizational Factors

Faheem Ahmed[1], Luiz Fernando Capretz[1], Shahbaz Ali Sheikh[2]

[1]Department of Electrical & Computer Engineering, Faculty of Engineering
[2]Management and Organizational Studies, Faculty of Social Science
University of Western Ontario, London, Ontario, Canada, N6A 5B9
fahmed@engga.uwo.ca, lcapretz@eng.uwo.ca, ssheik2@uwo.ca

**Abstract:** A good fit between the person and the organization is essential in a better organizational performance. This is even more crucial in case of institutionalization of a software product line practice within an organization. Employees' participation, organizational behavior and management contemplation play a vital role in successfully institutionalizing software product lines in a firm. Organizational dimension has been weighted as one of the critical dimensions in software product line theory and practice. A comprehensive empirical investigation to study the impact of some organizational factors on the performance of software product line practice is presented in this work. This is the first study to empirically investigate and demonstrate the relationships between some of the key organizational factors and software product line performance of an organization. The results of this investigation provide empirical evidence and further support the theoretical foundations that in order to institutionalize software product lines within an organization, organizational factors play an important role.

## I. INTRODUCTION

In recent years software development organizations have shown a growing interest in software product lines concept in order to compete in a global market that emphasizes cost, quality and adherence to the delivery schedule. Clement *et al.* [13] report that software product line engineering is a growing software engineering sub-discipline, and many organizations, including Philips®, Hewlett-Packard®, Nokia®, Raytheon®, and Cummins®, are using it to achieve extraordinary gains in productivity, time to market, and product quality. The economic potential of software product line has also been recognized in the software industry [7][64]. A software product line is a set of software intensive systems sharing a common, managed set of features that satisfy the specific needs of a particular market segment and are developed from a common set of core assets in a prescribed way [14]. There are other corresponding terminologies for software product line, which have been widely used in Europe, for example "product families", "product population", and "system families". In Europe, the acronym BAPO [65] (Business-Architecture-Process-Organization) is very popular for defining process concerns associated with software product lines. Business, architecture, process and organization are considered critical because they establish an infrastructure and manage the way the products resulting from software product lines make profits. The software product line strategy defines specific tasks for the organizational management, technical management, and software engineering aspects of product production [47].

Research has been reported [5][15][30][70] on the software product line process methodology including, product line architecture, commonality and variability management, core assets management, business case engineering, application and domain engineering, etc. However, the organizational aspects of software product lines such as organizational structure, roles and responsibilities, organizational learning, change management, conflict management, organizational culture, and organizational commitment are not discussed at large. This papers' main contribution is to increase the understanding of the influence of some of the key organizational factors by showing empirically that they play an imperative role in institutionalizing software product line within an organization. A quantitative survey of software organizations currently using the concept of software product lines over a wide range of operations, including consumer electronics, telecommunications, avionics, and information technology, was conducted and is reported in this paper. This survey is used to test the conceptual model and hypotheses of this study. The results of this study provide evidence that in order to institutionalize the concept of software product line into an organization, the management has to deal with many organizational factors.

## A. Institutionalization of Software Product Line

Institutionalization is the process by which a significantly new structure or practice is incorporated into a system of existing structures and practices [58]. Clements and Northrop [15] elaborate the institutionalization process of software product line in an organization from the perspective of product and core assets development activities. Institutionalizing a software product line from the perspective of product development anticipates the product development process as a routine and predictable activity in an organization in order to achieve the product line goals. Clements and Northrop [15] emphasize that institutionalizing a software product line from the perspective of managing and developing a core assets repository for software product line involves improving the processes that are associated with building, maintaining, and evolving the core assets. Those processes must become a part of standard organizational practice. In short, institutionalization of software product lines refers to the wide acceptance of the concept in the roots of the organization. It involves integrating or improving the processes within the organization that are associated with a product line infrastructure. Furthermore, introducing those processes as a part of the organizational character. The whole institutionalization process involves an organizational level culture and strong commitments to acquire knowledge, skills and motivations to effectively initiate, launch and manage software product lines. Institutionalization of software product lines requires that the concept have been entrenched at all levels of the organization. It must be supported with a necessary infrastructure of organization wide guidelines, required training, and resources.

Successful institutionalization of a software product line, in an organization has a profound impact on the product development behavior of the organization. It changes the mindset of the organization from single system development to a family of software products. The organizational theory focuses on the design and structure of the organization developing a software product line. The organizational behavior aims at understanding the behavior, attitude and performance of the people. The software product line requires enriching this concept within the roots of the overall organizational behavior. Organizational management plays a vital role in successfully institutionalizing software product line within an organization. It provides and coordinates the infrastructure required. Initiating and launching a software product line within an organization to gain benefits out of this approach is not sufficient alone. The alignment of organizational theory, management, and behavior are required in the process of institutionalization of software product line in an organization. Thus, organizational factors play a key role in institutionalizing software product lines within an organization.

## B. Software Product Line & Organizational Dimension: Related Work

The organizational dimension is perhaps the least addressed area in software product line research due to being a relatively new concept in software engineering paradigms. Much of the efforts have been spent on the process, architecture and business aspects of the software product line. Some scenarios of organizational structure for software product line are presented. The researchers generally highlight that domain-engineering unit and several application-engineering units are required from an organizational structure viewpoint. Bosch [6] presents four organizational models for software product lines: development department, business units, domain engineering units, and hierarchical domain engineering units. Bosch [6] also points out a number of factors that influence the organizational model such as geographical distribution, project management maturity, organizational culture and the type of systems. Macala *et al.* [43] report that software product line demands careful strategic planning, a mature development process, and the ability to overcome organizational resistance. Dikel *et al.* [23] share their experiences about initiating and maintaining software product lines at Nortel® and discuss organizational, management and staffing issues. The issues are grouped into a set of six organizational principles, which they believe are critical in the long-term success of a software product line. Jacobsen *et al.* [29] focus on roles and responsibilities of personnel within organizations dealing with software product lines. Mannion [44] elaborates that the management issues, organizational structure, culture and learning in context of successfully adopting the concept of software product line engineering needs close attention. Koh and Kim [37] conclude that all members of an organization must share their success stories and experience in order to successfully adopt the software product line approach.

Clements and Northrop [15] discuss organizational issues of software product line and identify four functional groups. The groups include the architecture group, the component-engineering group, the product line support group and the product development group. The organizational dimension of software product line deals with the way the organization is able to manage complex relationships and many employee responsibilities [64]. Toft *et al.* [63] propose "Owen molecule model" consist of three dimensions: organization, technology and business. The organizational dimension of the Owen molecule model deals with teams hierarchy, individual roles, operational models, individual interaction and communication etc. Introducing software product line practice into an organization significantly impacts the entire organization by fundamentally changing development practices, organizational structures, and task assignments [4]. Bayer *et al.* [2] at Fraunhofer Institute of Experimental Software Engineering (IESE) develop a methodology called PuLSE (Product Line Software Engineering) for the purpose of enabling the conception and deployment of software product lines within a large variety of enterprise contexts. PuLSE-BC is a technical component of PuLSE methodology. It deals with the ways to baseline organization and customized the PuLSE methodology to the specific needs of the organization. One of the support components of PuLSE is organization issue, which provide guidelines to set up and maintain the right organizational structure for developing and managing product lines. According to Birk *et al.* [4] introducing product line development to an organization can fundamentally change the development practices, organizational structures, and task assignments. These changes can in turn impact team collaboration and work satisfaction. Verlage and Kiesgen [67] report the case study of successful adoption of software product line and conclude that organizational structure and change management are significantly important areas of concern.

The summary of the related work presented in this sub-section exposes some key organizational factors such as organizational structure, organizational culture, conflict management, change management, organizational commitment and organizational learning. We used these key organizational factors as a set of independent variables in the empirical investigation presented in this paper in order to construct the research model of our investigation.

## II. ORGANIZATIONAL FACTORS: LITERATURE REVIEW OF CONCEPTS

The organizational theories help in learning how to better integrate people into engineering systems of all types. Researchers from various fields such as anthropology, economics, management, political science, psychology, scientific management, sociology and engineering have contributed towards the development of organizational theories. Organization is a planned coordination of activities of a number of people for the achievement of some common, explicit purpose or goal, through division of labor and function, and through a hierarchy of authority and responsibility [56]. Organizational theories, organizational behavior, and organizational management identified certain factors, which are termed as "organizational factors" in the rest of this paper. These organizational factors are considered to be the building blocks for integrating humans in cooperative working environment. The objective of this collaborative working environment is to achieve a common end goal in the presence of external and internal influences. The organizational factors used in this study are organizational structure, organizational culture, conflict management, change management, organizational commitment and organizational learning.

The organizational structure depicts the level of authority and the distribution of workload and responsibilities. Organizational theories provide guidelines to develop organizational structure in order to accomplish the goals of an organization. Champoux [11] elaborates that organizations are bounded systems. Wilson and Rosenfeld [71] define organizational structure as the established pattern of relationships between the parts of an organization. It outlines communication, control and authority pattern. According to Gordon [26] organizational structure refers to the delineation of jobs and reporting relationships in an organization. It coordinates the work behavior of employees in accomplishing the organizations' goals. Paterson [53] observes that success of any strategy depends heavily on its fit with the existing organizational structure. The structure of an organization is generally not a static phenomenon. The organization tends to change its structures under the circumstances of changing goals or technology.

The business, economic and technological events tend to force changes within the organization. The changes can be in strategy, organizational structure, process methodology, and technology or even in the organizational goals. The rapid and continual changes common to the present technological environment demand that organizations adopt changes through well-defined change management plans. Beckhard and Harris [3] consider organizational change as movement from the present state of the organization to some future or target state. Todd [62] defines change management as a structured and systematic approach, which provides a conceptual framework that encompasses strategy, politics, people and process. Cao *et al.* [9] observe that organizational change shows a diversity of the organization in its environment. It also shows the interaction of the technical and human activities within the organization. Introducing changes into the organization requires a comprehensive change management plan, which outlines the nature of the change and the procedures that will be used in the current working environment.

The organizational culture describes the overall organizational behavior and it helps in understanding the working environment of the organization. The organizational culture has a strong influence on the managerial and workforce behavior in carrying out day today tasks. Organizational culture has been characterized by many authors as a set of shared values, belief, assumptions, and practices that shape and guide employees' attitudes and behavior in the organization [39][51][72]. Organizational culture has something to do with the people and the unique quality and style of the organization [35]. Rosen [54] acknowledges that the internal orientation of employees is based primarily on the culture, beliefs, ethics and assumptions of the organizations' staff, and therefore, has the potential to be one of the most powerful influences on strategic management.

Organizational commitment is a work attitude that is directly related to employee participation and intentions to remain with an organization and is clearly linked to job performance [46]. Crewson [19] summarizes organizational commitment as a combination of three distinct factors with reference to employee participation: a strong belief and acceptance of the organizations' goals and values, eagerness to work hard for the organization, and a desire to remain a member of the organization. Schwepker [57] finds that stronger enforcements of ethical rules and codes are positively related to organizational commitment. The organizational commitment has profound impact on the job performance and job satisfaction of the employees. The job satisfaction allows employees to decide to remain with the organization for longer time.

In the software industry the technology is changing at an unprecedented rate of growth. Software organizations are more inclined towards learning new technology, methodology and processes in order to keep themselves up-to-date with the latest state of the art practice. Argyris [1] defines organizational learning as the process whereby members of the organization respond to changes in the internal and external environments of the organization by detecting errors, which they then correct so as to maintain the central features of the organization. Marquardt and Reynolds [45] define learning as a process by which individuals gain new knowledge and insights to change their behavior and actions. Hames [27] defines learning as encompassing the acquisition and practice of new methodologies, skills, attitudes, and values necessary to live in a world that is constantly changing. Lyles [42] observes that organizations do learn from their experiences and can remember incidents from the past that may influence future actions.

When people are interacting with each other, there is chance that they may have conflicts at different levels such as personal or task related. Walls and Callister [69] conclude conflict as a process in which one party perceives that its interests are being opposed or negatively affected by another party. Conflict management consists of diagnostic processes, interpersonal styles, negotiating strategies, and other interventions designed to avoid unnecessary conflict [38]. Hellriegel *et al.* [28] put forward four basic forms of conflict in an organization: goal, cognitive, affective, and procedural. Jehn [31] distinguishes between two kinds of intra-group conflict: task conflict and relationship conflict. Task conflict is a perception of disagreement among group members or individuals about the content of their decisions, and involves differences in viewpoints, ideas and opinions, whereas, relationship conflict is a perception of interpersonal incompatibility, and includes annoyance and animosity among individuals [48].

## III. RESEARCH MODEL AND HYPOTHESES OF THE STUDY

The main objective of the research model of this study is to analyze the association between organizational factors and the software product line performance. It is important to note here that although the literature of organizational theory and the theories of management have a long list of empirical investigations addressing organizational factors and their impact on the performance of an organization. This study is the first of its kind in the context of software product lines performance and key organizational factors at the best of our knowledge. This study provides an opportunity to empirically investigate the association between the key organizational factors and the software product line performance. The theoretical model to be empirically tested in this study is shown in Figure 1. The model examines the relationships of a number of independent variables arising from the concept of organizational theory, management, and organizational behavior on the dependent variable of software product line performance within an organization. The main objective of this study is to investigate the answer to the following research question:

**Research Question:** What is the impact of organizational factors on the overall performance of a software product line?

There are six independent and one dependent variable in this research model. The six independent variables are called "organizational factors" in the rest of this paper. They include organizational structure, change management, organizational culture, conflict management, organizational commitment and organizational learning. The dependent variable of this study is the software product line performance of an organization. The multiple linear regression equation of the model is depicted by Equation (I)

$$\text{Software Product Line Performance} = \beta_0 + \beta_1 f_1 + \beta_2 f_2 + \beta_3 f_3 + \beta_4 f_4 + \beta_5 f_5 + \beta_6 f_6 \quad \text{-----------(I)}$$

Where $\beta_0, \beta_1, \beta_2, \beta_3, \beta_4, \beta_5, \beta_6,$ are coefficients, and $f_1, f_2, f_3, f_4, f_5, f_6,$ are the six independent variables. In order to empirically investigate the research question we hypothesize the following:

**H1:**   Organizational structure has a positive impact on software product line performance.

**H2:** Effective Change management planning, execution, and control have a positive impact on software product line performance.

**H3:** Organizational culture is positively associated with the performance of a software product line.

**H4:** Organizational commitment plays a positive role in software product line performance.

**H5:** Organizational learning is positively associated with software product line performance.

**H6:** The performance of a software product line is positively associated with effective conflict management within an organization.

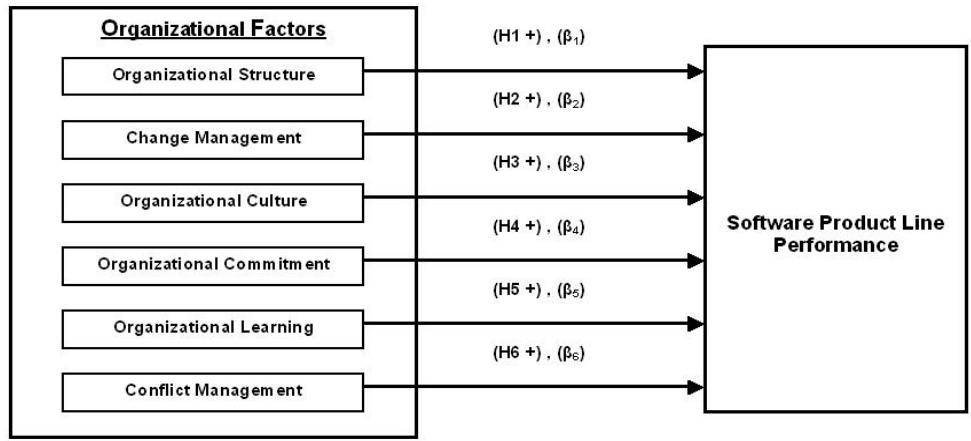

**Figure 1: Research Model**

## IV. RESEARCH METHODOLOGY

Software organizations covering a wide range of operations, such as consumer electronics, telecommunication, avionics, automobiles, and information technology, and having been involved in using the software product line approach for more than three years were the target population for this study. Initially we placed our request to the software product line research community for their participatation in this study at the discussion forum of the software product line website, maintained by www.softwareproductlines.com. We also sent personalized emails to different organizations. Nine organizations agreed to participate in this study with the mutual understanding to keep the names of the organizations confidential. The participating organizations are North American and European multinational companies. The organizations differed in size and range from medium to large-scale. We assume that the medium scale organization has number of employees around 2000 to 3000, whereas a large-scale organization has more then 3000 employees. It is important to note here that the size of the organization in terms of number of employees is based on total number of employees in the organization working in various departments.

We requested the organizations under study to distribute the questionnaire within various departments. The respondents, on average, had been associated with the organizations for the last three years. The minimum qualification of respondents was an undergraduate university degree and the maximum was a Ph.D. degree. The respondents generally belonged to middle and senior management and development categories. They either had roles in making policy or in implementing organizational strategies from top to bottom. We received a minimum of three and a maximum of six responses from each organization. The total respondents were 40 altogether.

### A. Measuring Instrument

In this study, we collected data on the organizational factors and the perceived level of software product line performance identified in the research model shown in Figure 1. The questionnaire presented in Appendix-I was used to serve as a source of first contact in learning two objectives. First, up to what extent the organizational factors were practiced within each organization dealing in the software product line. Secondly, the perceived level of organizational performance in software product line concept. The questionnaire required respondents to indicate the extent of their agreement or disagreement with statements using a five-point Likert scale. We used twenty-four separate items to measure the independent variables. We used four items to measure each organization's performance in the software product lines. Previous researches on organizational theory, organizational management and organizational behavior were reviewed to ensure that a comprehensive list of measures were included in constructs of organizational factors. In order to measure the extent to which each of the six organizational factors was practiced in organizations we used multi-item, five-points Likert scale. The Likert scale ranged from "strongly disagree" (1) to "strongly agree" (5) for all items associated with each variable. Four items for each independent variable were designed to collect measures on the extent to which the variable is practiced within each organization. The items for all six organizational factors are labeled sequentially in Appendix-I and are numbered 1 through 24.

We measured the dependent variable, i.e. software product line performance for the past three years, with respect to cost and development time reductions, market growth, and financial strengths based on the multi-item, five-point Likert scale. The items were specifically designed for collecting measures for this variable and are labeled sequentially from 1 through 4 in Appendix-I. To the best of our knowledge, this is the first study of its kind within the area of software product lines. Therefore all items shown in Appendix-I are written specifically for this empirical investigation.

## B. Reliability and Validity Analysis of Measuring Instrument

Reliability and validity are two integral features of empirical studies. Reliability refers to the reproducibility of a measurement, whereas validity refers to the agreement between the value of a measurement and its true value. We conducted reliability and validity analysis of the measuring instruments designed specifically for this empirical investigation by using the most common approaches generally used in empirical studies. The reliability of the multiple-item measurement scales of the six organizational factors was evaluated by using internal-consistency analysis. Internal-consistency analysis was performed using coefficient alpha [20]. Table-I reports the reliability analysis, the coefficient alpha ranges from 0.72 to 0.88. Nunnally and Bernste [49] found that a reliability coefficient of 0.70 or higher for a measuring instrument is satisfactory. The other reliability literature such as *van de* Van and Ferry [66] suggests that a reliability coefficient of 0.55 or higher is satisfactory, and Osterhof [52] concluded that 0.60 or higher is satisfactory. Therefore we determined that all variable items developed for this empirical investigation were reliable.

We observed the content validity of the items included in each organizational factor, following the general recommendations of Cronbach [21] and Straub [61], by carrying out a comprehensive literature survey to include possible items in the variable scales. We also held discussions with the representatives of the organizations to finalize the proposed independent variables and items included in each variable. Statistical and psychological-testing experts reviewed the proposed scales. We conducted pilot tests, which led to modifications in the variable items, based on the suggestions of respondents, which improved the content validity.

**Table-I: Coefficient Alpha & Principal Component Analysis of Variables**

| Organizational Factors | Item No. | Coefficient α | PCA Eigen-Value |
|---|---|---|---|
| Organizational Structure | 1-4 | 0.79 | 2.51 |
| Change Management | 5-8 | 0.86 | 2.86 |
| Organizational Culture | 9-12 | 0.72 | 2.23 |
| Organizational Commitment | 13-16 | 0.78 | 2.51 |
| Organizational Learning | 17-20 | 0.88 | 3.01 |
| Conflict Management | 21-24 | 0.73 | 2.38 |

Convergent validity, according to Campbell and Fiske [8] occurs when the scale items in a given construct move in the same direction and thus highly correlate. The principal component analysis [18] performed and reported for all six organizational factors in Table-I provides a measure of convergent validity. We used eigen value [33] and scree plot [10] as reference points to observe the construct validity using principal component analysis. In this study, we used eigen value-one-criterion, also known as Kaiser criterion [34] [60], which means any component having an eigen value greater then one was retained. Eigen value analysis revealed that five out of six variables completely formed a single factor, whereas in the case of organizational structure two components are formed. The eigen value for the second component is slightly higher than the threshold of 1.0. The scree plots clearly showed a cut-off at the first component. Therefore, the convergent validity can be regarded as sufficient.

We used multiple regression analysis to determine the criterion validity of the six organizational factors and software product line performance. Organizational factors were used as predictor variables and software product line performance was used as a criterion variable. The multiple correlation coefficient observed, was 0.85. Cohen [17] concluded that a multiple correlation coefficient higher than 0.51 corresponds to a large effect size. Therefore, we observed the criterion validity of the variables to be sufficient. The measurements of reliability and validity analysis showed that the measurement procedures used in this study had the required level of psychometric properties.

## C. Data Analysis Techniques

To analyze the research model and check the significance of hypotheses H1–H6, we used various statistical analysis techniques. Initially we divided the data analysis activity into three phases. Phase-I dealt with normal distribution tests and parametric statistics. Phase-II dealt with non-parametric statistics. In order to reduce the threats to external validity due to small sample size, we used both statistical approaches of parametric and non-parametric methods. We tested for the normal distribution of all the organizational factors using mean, standard deviation, kurtosis and skewness techniques, and found the values for all these tests to be within the acceptable range for the normal distribution with some exceptions. We made some modification to the data received from respondents before performing statistical analysis. Since all of the six independent variables and the dependent variable's measuring instrument had multiple items. Therefore we added their ratings to obtain a composite score for that measure before performing statistical analysis. The statistical analysis results reported in this paper are based on data received from all the respondents.

We conducted tests for hypotheses H1–H6 using parametric statistics, such as the Pearson correlation coefficient and one tailed t–test in Phase-I. In Phase-II of non-parametric statistics, we conducted tests for hypotheses H1–H6 using the Spearman correlation coefficient. Phase-III dealt with testing the hypotheses of the research model of this study using the technique of Partial Least Square (PLS). The PLS technique helps when complexity, non-normal distribution, low theoretical information, and small sample size are issues [25] [32]. Since small sample size was one of the major limitations in this study, therefore, we used the PLS technique to increase the reliability of the results as well. One of the main reasons for small sample size is that software product line is a relatively young concept in the software industry. Not many organizations have been dealing in software product lines over the last three years. The statistical calculations were performed using Minitab® 14 software.

## V. DATA ANALYSIS AND RESULTS

### A. Hypotheses Testing Phase-I

We examined the Pearson correlation coefficient and t-test between individual independent variables (organizational factors) and the dependent variable (software product line performance) of the research model shown in Figure 1 in order to test hypotheses H1–H6. The result of the statistical calculations for the Pearson correlation coefficient is reported in Table-II. The Pearson correlation coefficient between organizational structure and software product line performance was positive (0.76) at P < 0.01, and thus provided a justification to accept the H1 hypothesis. The hypothesis H2 was accepted based on the Pearson correlation coefficient (0.74) at P < 0.01, between change management and performance of software product line. The correlation coefficient of 0.61 at P < 0.01 was observed between the software product line performance and organizational culture. The positive correlation coefficient of 0.76 at P < 0.01 meant that H4 was accepted. Hypothesis H5 was accepted after analyzing the Pearson correlation coefficient (0.78 at P < 0.01). The hypotheses H6 (correlation: 0.28) was not found significant at P < 0.05. Therefore the hypotheses H6 that deals with conflict management and software product line performance was rejected. Hence, it was observed and is reported here that hypotheses H1, H2, H3, H4, and H5, were found statistically significant and were accepted whereas H6 was not supported and was therefore rejected.

**Table-II: Hypotheses Testing Using Parametric and Non-Parametric Correlation Coefficients**

| Hypothesis | Organizational Factors | Pearson Correlation Coefficient | Spearman Correlation Coefficient |
|---|---|---|---|
| H1 | Organizational Structure | 0.76* | 0.73* |
| H2 | Change Management | 0.74* | 0.71* |
| H3 | Organizational Culture | 0.61* | 0.58* |
| H4 | Organizational Commitment | 0.76* | 0.80* |
| H5 | Organizational Learning | 0.78* | 0.79* |
| H6 | Conflict Management | 0.28** | 0.30** |

* Significant at P < 0.01     ** Insignificant at P > 0.05

## B. Hypotheses Testing Phase-II

In Phase-II we conducted non-parametric statistical technique using Spearman correlation coefficient to test the hypotheses H1–H6. Table-II also reports the observation made in this testing phase. Hypothesis H1 was statistically significant at $P < 0.01$ with Spearman correlation coefficient of 0.73. A positive association was observed between change management and software product line performance (Spearman: 0.71 at $P < 0.01$). H3, which deals with organizational culture and software product line performance, was accepted (Spearman: 0.58 at $P < 0.01$. The Spearman correlation of (0.80 at $P < 0.01$) was observed for H4. A positive Spearman correlation of 0.79 at $P < 0.01$ resulted in accepting H5. The hypothesis H6 between conflict management and software product line performance was found statistically insignificant because the observed P-value was greater then 0.5. Hence, it was observed and is reported here that hypotheses H1, H2, H3, H4, and H5, were found statistically significant and were accepted whereas H6 was not supported and was therefore rejected.

## C. Hypotheses Testing Phase-III

In Phase-III of hypotheses testing, we used the PLS technique to overcome some of the associated limitations and to cross validate with the results observed using the approaches of Phase-I and Phase-II. We tested the hypothesized relationships, i.e. H1–H6, by examining their direction and significance. In PLS we placed software product line performance as the response variable and individual organizational factor as the predicate. Table-III reports the results of the structural tests of the hypotheses. It contains observed values of path coefficient, $R^2$ and F-ratio. The path coefficient of organizational structure was found to be 0.68, $R^2$: 0.58 and F-ratio (53.70) was significant at $P < 0.01$. Change management had positive path coefficient of 0.50 with $R^2$: 0.54 and at $P < 0.01$ F-ratio was 45.87. Organizational culture (Path coefficient: 0.58, $R^2$: 0.37, F-ratio: 22.49 at $P < 0.01$) had the same direction as proposed. Organizational commitment (Path coefficient: 0.70, $R^2$: 0.59, F-ratio: 54.90 at $P < 0.01$) also had the same direction as proposed in H4. Organizational learning (Path coefficient: 0.54, $R^2$: 0.61, F-ratio: 61.21 at $P < 0.01$) was also found in accordance with H5. Conflict management had path coefficient of 0.32 at a very low $R^2$ of 0.08 and F-ratio of 3.38 was not found significant at $P < 0.05$. All in all, the hypotheses H1, H2, H3, H4, and H5 showed significant at $P < 0.01$ with a positive path coefficient and were in the same direction as proposed. The hypotheses H6 that deals with conflict management and software product line performance was not found to be statistically significant at $P < 0.05$.

**Table-III: Hypotheses Testing Using Partial Least Square Regression**

| Hypothesis | Organizational Factors | Path Coefficient | $R^2$ | F-Ratio |
|---|---|---|---|---|
| H1 | Organizational Structure | 0.68 | 0.58 | 53.70* |
| H2 | Change Management | 0.50 | 0.54 | 45.87* |
| H3 | Organizational Culture | 0.58 | 0.37 | 22.49* |
| H4 | Organizational Commitment | 0.70 | 0.59 | 54.90* |
| H5 | Organizational Learning | 0.54 | 0.61 | 61.21* |
| H6 | Conflict Management | 0.32 | 0.08 | 3.38** |

\* Significant at $P < 0.01$    \*\* Insignificant at $P > 0.05$

## D. Testing of Research Model

The linear regression equation of the research model is illustrated by Equation (I). The purpose of research model testing was to provide empirical evidence that organizational factors play a considerable role in software product line performance. The testing process consists of conducting regression analysis and reporting the values of the model coefficients and their direction of association. We placed software product line performance as response variable and individual organizational factor as predicators. The analysis also reports the results of two-tailed t-tests conducted and their statistical significance. Table-IV reports the regression analysis of the research model. The path coefficient of five out of six variables: organizational structure, change management, organizational culture, organizational commitment and organizational learning were found positive and their t-statistics was also observed statistically significant at either $P < 0.01$ or $P < 0.05$. The path coefficient of conflict management was found negative. Negative t-statistics and $P > 0.05$ make conflict management statistically insignificant in this research model. The adjusted $R^2$ of overall research model was observed 0.85 with a F-ratio of 37.71 significant at $P < 0.01$.

**Table-IV: Linear Regression Analysis of Research Model**

| Model Coefficient Name | Model Coefficient | Coefficient Value | t-value |
|---|---|---|---|
| Organizational Structure | $\beta_1$ | 0.18 | 2.21** |
| Change Management | $\beta_2$ | 0.14 | 2.23** |
| Organizational Culture | $\beta_3$ | 0.23 | 3.17* |
| Organizational Commitment | $\beta_4$ | 0.20 | 2.43** |
| Organizational Learning | $\beta_5$ | 0.22 | 3.14* |
| Conflict Management | $\beta_6$ | -0.08 | -1.05*** |
| Constant | $\beta_0$ | 1.47 | 1.14 |
| $R^2$ | | 0.87 | Adjusted $R^2$ = 0.85 |
| F-Ratio | | 37.71* | |

\* Significant at P < 0.01     ** Significant at P < 0.05     ***Insignificant at P > 0.05

## VI. SENSITIVITY ANALYSIS OF THE STUDY

Empirical investigations are subject to a number of limitations, which may result in threats to external validity and reliability. Results from an empirical study depend on data, the statistical model and the statistical techniques used. Sensitivity analysis is defined as the investigation of how research model misspecification and anomalous data points influence results [50]. According to Kitchenham *et al.* [36], in empirical investigations, it is important to perform a sensitivity analysis to understand how individual data points or clusters of data relate to the behavior of the whole collection. Saltelli *et al.* [55] define sensitivity analysis as the study of how the variation in the output of a model can be apportioned, qualitatively or quantitatively, among model inputs. The sensitivity analysis allows the researchers to understand how the research model behaves on changing inputs. It further supports the empirical investigations in terms of reliability and validity.  In this section we reported two sensitivity analysis tests. The first test deals with an inter-rater agreement because we have a varying number of respondents within the same organization. The second test performs sensitivity analysis of the overall research model shown in Figure 1.

**Table-V: Inter-Rater Agreement Analysis**

| Organization | Kendall Statistics | | Kappa Statistics | |
|---|---|---|---|---|
| | Kendall's Coefficient Of Concordance (W) | $\chi^2$ | Kappa Coefficient | Z |
| A | 0.47 | 65.72* | 0.20 | 5.07* |
| B | 0.60 | 82.94* | 0.45 | 10.53* |
| C | 0.51 | 35.58** | 0.21 | 1.82** |
| D | 0.70 | 48.86* | 0.46 | 4.67* |
| E | 0.44 | 30.94*** | 0.16 | 0.70*** |
| F | 0.49 | 57.29* | 0.28 | 4.78* |
| G | 0.53 | 61.06* | 0.33 | 5.34* |
| H | 0.43 | 50.03* | 0.23 | 4.57* |
| I | 0.63 | 58.22* | 0.34 | 5.40* |

\* Significant at P < 0.01     ** Significant at P < 0.05     ***Insignificant at P > 0.05

### A. Inter-Rater Agreement Analysis

Inter-rater agreement corresponds to reproducibility in the evaluation of the same process according to the same evaluation specification [41]. According to El Emam [24] the inter-rater agreement is concerned with the extent of agreement in the ratings given by independent assessors to the same software engineering practices. We received a varying number of responses from different organizations. A varying number of respondents within an organization may have conflicting opinions about the performance of the software sproduct line. Therefore, there is a need to perform an inter-rater agreement analysis to provide information about the extent of agreement among the raters within one organization. The Kendall coefficient of concordance (*W*) [68] is often preferred to evaluate inter-rater agreement in comparison to other methods such as Cohen's Kappa [16] in case there is ordinal data. "*W*" is an index of the divergence of the actual agreement shown in the data from the possible perfect agreement.

In order to ensure the reliability and validity of this empirical investigation, we conducted and reported the inter-rater agreement analysis using Kendall's and Kappa statistics. Table-V reports the Kendall and Kappa statistics of the nine organizations that participated in this study. Values of Kendall's *W* and Kappa coefficient can range from 0 to 1, with 0 indicating perfect disagreement, and 1 indicating perfect agreement [40].

## B. Sensitivity Analysis of the Research Model

The research model shown in Figure 1 consists of six-indepdent variables termed as organizational factors in this study. The results of the empirical investigation reported in Section IV shows that five out of these six independent variables are positively associated with the dependent variable of software product line performance. Conflict management did not show any significant relationship with the performance of software product line. The objectives of the sensitivity analysis of research model are twofold. First, it studies the impact of each independent variable on the overall output of the model. Secondly, it learns the impact of conflict management on the overall output of the model because it has not been supported by the study. We used Fourier Amplitude Sensitivity Test (FAST) and Sobol methods to conduct and report sensitivity analysis of the research model of this empirical investigation. The FAST method is commonly used to estimate the ratio of the contribution of each input to the output variance with respect to the total variance of the output as the first order sensitivity index. FAST can identify the contribution of individual inputs to the expected value of the output variance [22]. FAST does not assume a specific functional relationship such as linear or monotonic in the model structure, and thus works for both monotonic and non-monotonic models [55].

The method of Sobol [59] apportions the output variance among individual inputs and their interactions. The method of Sobol can cope with both non-linear and non-monotonic models, and provides a quantitative ranking of inputs [12]. Sobol's method provides insight with respect to the main effect, interaction effect, and total effect of each input with respect to the output of interest. The main effect of each input represents the fractional unique linear contribution of the input to the output variance. The sensitivity analysis of the research model is reported in Table VI. It illustrates that all the five independent variables, which were positively associated with the dependent variable of the research model, contribute significantly to the output of the model. Conflict management, which was not significantly associated with the performance of software product line, has very low contribution (FAST: 4%, Sobol: 3%) in the overall output of the model. The sensitivity evaluation calculations were performed using SimLab 2.2 software.

**Table-VI: Sensitivity Analysis of the Research Model**

| Organizational Factors | FAST Sensitivity Index (%) | Sobol Sensitivity Index (%) |
|---|---|---|
| Organizational Structure | 15 | 10 |
| Change Management | 13 | 12 |
| Organizational Culture | 15 | 19 |
| Organizational Commitment | 20 | 12 |
| Organizational Learning | 37 | 29 |
| Conflict Management | 4 | 3 |

## VII. DISCUSSION

The software product line is an inter-disciplinary concept, which has its roots in software engineering, business, management and organizational sciences. This research enables organizations to understand the effectiveness of the relationships and interdependency of organizational factors and the software product line process. This study provides an opportunity to empirically investigate the association between the organizational factors and software product line performance. The organizational factors we employed were divided into two broad categories: organizational management and organizational behavior. The results provide the first empirical evidence and support for the theoretical foundations that organizational factors play a critical role in the institutionalization of software product line within an organization. The organization in the business of software product line has to deal with multiple organizational factors in addition to their efforts in software development. This results in institutionalizing software product line in an organization, which in turn has the potential to achieve maximum benefits out of this approach.

Organizational structure relates to how different parts of an organization work in a setting that helps in achieving the overall goals of the organization. This study finds a positive association between organizational structure and software product line performance. Software product line requires setting up an internal structure of the organization and other supporting mechanism such as coordination and communication. The organizational theory provides various organizational structures that had been successfully employed in various settings. The concept of software product line envisages a structure of overlapping processes rather then fixed and static ones. The theoretical foundations of this concept divide the overall engineering process into two broad areas of application and domain engineering and foresee a strong coordination and communication between them. The identification of roles and mapping of the roles to the engineering processes requires management contemplation. Verlage and Kiesgen [67] report that the roles and mapping of the roles to the processes are not fixed rather then they are interchangeable or more precisely dynamic, while presenting case study of the successful implementation of software product lines in their organization. The organizations that have well defined structures spread over clearly identified roles for individual employees along with strong coordination and communication are more likely to institutionalize software product line as compared to organizations that have structures that do not support close coordination and communication.

We found a positive impact of better change management on the performance of software product line in our empirical investigation. Introducing a new practice such as product line is relatively difficult in the existing setup of an organization if it is not being introduced with a proper change management plan. Even the best strategy is bound to fail if there is a consistent resistance to innovation and new technology from within the organization. The successful implementation of any process methodology ultimately depends on how employees perceive that change. A certain degree of resistance is quite normal when a new technology is introduced in an organization. However, this resistance could go away if employees understand that this change is positive and is in the best interests of the organization as well as for them. An effective change management strategy partly hinges on how the strategy is communicated to the people who are responsible for its implementation. Moving from single product development to a line of products is a significant change. Thus, the organizations that communicate the importance of this change via clear guidelines, with a road map on how to adapt to this change can be more successful in institutionalizing software product lines.

Organizational culture is another determinant in the success of software product lines. In this empirical investigation we found a positive association between organizational culture and the performance of software product line. Organizational culture refers to the norms and unwritten ways of working in an organization. There are organizational cultures that do not accept any new idea and resist to any change. The culture is a reflection of people's commitment to the credo of the organization. It reflects the nature of relationships between management and employees. If an organization has a hierarchical culture where all the decisions are made from the top and dictated to the bottom, employees normally have the tendency to resist. On the other hand, if an organization encourages innovation and new thinking then the employees can express their opinions and suggestions in an open environment. Any new idea with an expected positive impact on organizational goals is readily assimilated into the current process without much resistance. This results in overall conducive working culture of the organization. The software product line requires a culture of openness where employees have the chance to participate in discussions and has the power to express their views. An organizational culture that supports teamwork, sharing of experiences, innovation and learning has the high potential to institutionalize software product line into the firm. Particularly, an organization with a culture that supports the reusability of software assets is more likely to succeed in moving from single product development to a systematic line of products.

The findings of this empirical investigation confirm a positive relationship between organizational commitment and software product line performance. The commitment of an organization to initiate and adopt a particular practice or methodology requires sufficient actions to be well communicated across the organization. A software product line requires the initial cost of setup and the pay back period is relatively longer than the single product development approach. This transitional period requires strong commitment from individuals, groups and management at all levels to adopt software product line concept and reassure themselves at various times. The organizational policies such as business vision and strategic planning must highlight the development of software product line as top priority in order to reflect the organizational commitments. These policies must be well communicated to the employees so that they understand the significance of this approach in achieving the organizational goals. The success of any long-term strategy in an organization begs the commitments of its employees as well. The management must create a good working environment including: well-defined job placement, promotion strategy, appreciation and reward system, job security, competitive compensation etc. in order to increase the commitments of the employees with the organization. Increasing the knowledge of employees through regular training sessions helps them to understand their job, which in turn increases their commitments to perform.

Organizational learning is one of the keys to success. This empirical investigation finds a positive impact of organizational learning on the overall performance of software product line. In the current global worldview, where technology is changing at an unpredictable rate, regular training and continuous learning can provide possible competitive edge. Most of the resistance to change comes from ignorance and lack of knowledge. If employees are provided with training and there is a culture that rewards learning, people will extend their full support to the demands of every new strategy. In the context of software product line, a successful institutionalization requires an organization wide learning culture that helps to initiate, launch and maintain product line. Organizational learning includes both employees and organization itself. In case of software product line, we can classify organizational learning into two domains: external and internal. External learning of the organization captures necessary knowledge about customers, competitors, external environment and market segments. These make the best use of product line strategy by exploiting the product characteristics. This part of learning helps an organization to capture major market share. Internal learning on the other hand requires acquiring, transferring and sharing of software product line methodology, ideas of process improvement and the understanding of the cross functional requirements of product lines in individuals, groups and the organization. Learning is a continuous process especially for organizations that attempt to institutionalize software product lines.

An organization is a systematic structure of rules, policies and processes. The successful implementation of any policy in general and software product line in particular depends on how organization manages personal and organizational conflicts. Personal conflicts may hinder the development and implementation of long-term strategies. If conflicts are not handled in a healthy environment, they may affect employees' morale and productivity. The presence of a clear-cut organizational policy about the conflict resolution provides an impersonal touch to personal conflicts. All conflicts are resolved in a professional environment that has least impact on employees' morale and productivity. At the same time, positive, impersonal conflicts are very important for the growth of the organization. Any organization that stifles healthy debate and professional conflicts may find itself locked in a box and stuck with stagnant growth. The institutionalization of software product line requires stable and conflict free environment. In this study at an early stage of research model development we expected that the organization that have well defined personal conflict resolution policies and encourages positive, impersonal and professional conflicts would be more successful in terms of the performance of software product line.

The findings of this empirically investigation do not statistically provide a significant support to the positive association of conflict management and software product line performance. Although the direction of association between the performance of software product line and conflict management was found to be negative in this study but the result was not supported by a significant statistical level of confidence. Therefore it is concluded that this study is not able to find an answer about the association and impact of conflict management and software product line performance.

## A. Limitations of the Study & Threats to External Validity

Empirical investigations are subject to certain limitations. That is the case with this study. The first limitation is the choice and selection of independent variables in this study. We used six independent variables to relate with the dependent variable of software product line performance. There may be other organizational factors that influence the performance of software product lines in addition to these six but we kept the scope of this study within organizational management and behavior. Some other contributing factors to performance of software product lines, such as: organization size, economic, experience in software development and political conditions are not considered in this study. We concentrated only on the organizational factors.

The second observable limitation of this study is small sample size. The software product line is a relatively young concept in software development, and not many of the organizations in the software industry have institutionalized and launched this concept. Therefore collecting data from the software industry was a limitation, which leads to small sample size. The small sample size in terms of number of organizations and respondents has a potential threat to the external validity of this study. The one major reason behind small number of participating organizations is our initial criteria set of three years of experience in software product line development. There are not many organizations having the required level of experience in the business of software product line in particular due to relatively young age of this concept. The reason behind choosing the three years experience in software product line, as a criteria set is the characteristics of long-term payback period of software product line development. In order to enhance the external validity, we intended to ensure that organizations have started enjoying the benefits of software product line in terms of pay back or at least some potential benefits are apparent now. The number of respondents from organizations was beyond our control as we requested at the organizational level to distribute the survey and provide us feedback.

The third notable limitation of this study is bias in decision-making. Although we used multiple respondents within the same organization to reduce bias, bias still is a core issue. We asked the respondents to consult major sources of data at their organization, i.e., documents, plans, models, and actors before responding to a particular item in order to reduce the human tendency to over- or under-estimate when filling in questionnaires. The items were designed using accepted psychometric principles, but the measurement is still largely based on the subjective assessment of an individual. Besides its general and specific limitations, this study contributes significantly in the area of software product lines and helps to understand the organizational dimension of software product lines.

## VIII. CONCLUSION & FUTURE WORK

Software engineering, business, management and organizational sciences provide foundations for the concept of software product line. It has turned to be an inter-disciplinary concept. This research facilitates better understanding of the organizational dimension of software product line. Our main objective was to empirically investigate the effect of organizational factors on the performance of software product line and find answer to the research question put forward in this investigation. Results of this empirical investigation demonstrate that organizational factors facilitate better software product line performance. Empirical results of this study strongly support the hypotheses that organizational structure, culture, commitment, learning and change management, are positively associated with the performance of software product line in an organization. We did not find any significant statistical support for conflict management and are unable to find an answer about the association and impact of conflict management on software product line performance. The study conducted and reported here is the first of its kind in the area of software product lines. This research will enable organizations to better understand the effectiveness of the relationships of organizational factors and software product line. This research also reinforces current perceptions about the significance of organizational factors and their impact on successful institutionalization of software product line. The organizations in the business of software product lines need to take into consideration multiple key organizational factors over and above their efforts to develop software in order to institutionalize this concept.

Currently, we are working on developing a *Process Maturity Model* for process assessment of software product line. This work has provided the empirical justification to include these organizational factors in evaluating the organizational dimension of software product line process maturity.

## REFERENCES

[1] C. Argyris, Double-loop learning in organizations, Harvard Business Review 55 (1977) 115-125.

[2] J. Bayer, O. Flege, P. Knauber, R. Laqua, D. Muthig, K. Schmid, T. Widen and J.M. DeBaud, PuLSE: a methodology to develop software product lines, in: Proceedings of the 5th ACM SIGSOFT Symposium on Software Reusability, 1999, pp. 122-131.

[3] R. Beckhard, and R.T. Harris, Organizational transitions: managing complex change, Addison-Wesley, 1987.

[4] G. H Birk, I. John, K. Schmid, T. von der Massen and K. Muller, Product line engineering, the state of the practice, IEEE Software 20 (6) (2003) 52-60.

[5] J. Bosch, Design and use of software architectures: adopting and evolving a product-line approach, Addison Wesley, 2000.

[6] J. Bosch, Software product lines: organizational alternatives, in: Proceedings of the 23rd International Conference on Software Engineering, 2001, pp. 91-100.

[7] G. Buckle, P.C. Clements, J.D. McGregor, D. Muthig, and K. Schmid, Calculating ROI for Software Product Lines, IEEE Software, 21 (3) (2004) 23-31.

[8] D.T. Campbell and D.W. Fiske, Convergent and discriminant validation by the multi-trait multi-method matrix, Psychological Bulletin 56 (2) (1959) 81-105.

[9] G. Cao, S. Clarke, and B. Lehaney, A systematic view of organizational change and TQM, The TQM Magazine 12 (3) (2000) 186-93.

[10] R.B. Cattell, The scree test for the number of factors, Multivariate Behavioral Research 1 (1966) 245-276.

[11] J. E. Champoux, Organizational behavior: essential tenets for a new millennium, Southwestern College Publishing, 2000.

[12] K. Chan, S. Tarantola, A. Saltelli, and I.M. Sobol, Variance-based methods in sensitivity analysis. John Wiley and Sons, New York, 2000.

[13] P.C. Clements, L.G. Jones, L.M. Northrop and J.D. McGregor, Project management in a software product line organization, IEEE Software 22 (5) (2005) 54-62.

[14] P.C. Clements, On the importance of product line scope, in: Proceedings of the 4th International Workshop on Software Product Family Engineering, 2001, pp. 69-77.

[15] P. C. Clements, and L.M Northrop, Software product lines practices and pattern, Addison Wesley, 2002.

[16] J. Cohen, A coefficient of agreement for nominal scales, Educational and Psychological Measurement 20 (1960) 37-46.

[17] J. Cohen, Statistical power analysis for the behavioral sciences, second ed. Hillsdale, N.J. 1988.

[18] A.L. Comrey, and H.B. Lee, A first course on factor analysis, second ed. Hillsdale, 1992.

[19] P. Crewson, Public service motivation: building empirical evidence of incidence and effect, Journal of Public Administration Research and Theory 7 (1997) 499-518.

[20] L.J. Cronbach, Coefficient alpha and the internal consistency of tests, Psychometrica, 16 (1951) 297-334.

[21] L.J. Cronbach, Test validation, Educational Measurement, (1971) 443-507.

[22] R.I. Cukier, H.B. Levine, and K.E. Shuler, Nonlinear sensitivity analysis of multi-parameter model systems, Journal of Computational Physics 26 (1) (1978) 1-42.

[23] D. Dikel, D. Kane, S. Ornburn, W. Loftus and J. Wilson, Applying software product-line architecture, IEEE Computer 30 (8) (1997) 49-55.

[24] K.El Emam, Benchmarking kappa: inter-rater agreement in software process assessments, Empirical Software Engineering 4 (2) (1999) 113-133.

[25] C. Fornell, and F. L. Bookstein, Two structural equation models: LISREL and PLS applied to consumer exit voice theory, Journal of Marketing Research 19 (1982) 440-452.

[26] J.R. Gordon, Organizational Behavior: A diagnostic approach, Prentice Hall, New Jersey, 2002.

[27] R. D. Hames, The management myth, Business and Professional Publishing, Sydney, 1994.

[28] D. Hellriegel, J.W. Jr. Slocum, R.W. Woodman and N.S. Bruning, Organizational behavior, ITP Nelson, Canada, 1998.

[29] I. Jacobsen, M. Griss and P. Jonsson, Software reuse - architecture, process and organization for business success, Addison Wesley, 1997.

[30] M. Jazayeri, A. Ran, and F. *van der* Linden, Software architecture for product families: principles and practice, Addison Wesley, 2000.

[31] K.A. Jehn, A multi-method examination of the benefits and detriments of intra-group conflict, Administrative Science Quarterly 40 (1995) 256-82.

[32] K. Joreskog and H. Wold, Systems under indirect observation: causality, structure and prediction, The Netherlands: North Holland, 1982.

[33] H.F. Kaiser, A second generation little jiffy, Psychometrika 35 (1970) 401-417.

*Journal of Systems and Software, Volume 80, Issue 6, pp. 836-849, Elsevier, June 2007*
*DOI: https://doi.org/10.1016/j.jss.2006.09.010*[34] H.F. Kaiser, The application of electronic computers to factor analysis, Educational and Psychological Measurement 20 (1960) 141-151.

[35] R.H. Kilmann, M.J. Saxton, and R. Serpa, Gaining control of the corporate culture, Jossey-Bass, San Francisco, CA, 1985.

[36] B. A. Kitchenham, S. L. Pfleeger, L. M. Pickard, P. W. Jones, D. C. Hoaglin, K. El Emam and J. Rosenberg, Preliminary guidelines for empirical research in software engineering, IEEE Transactions on Software Engineering 28 (8) (2002) 721-734.

[37] E. Koh and S. Kim, Issues on adopting software product line, in: Proceedings of the 11[th] Asia-Pacific Conference on Software Engineering, 2004, pp. 589.

[38] J. Kottler, Beyond blame: A new way of resolving conflicts in relationships, Jossey-Bass, San Francisco, 1994.

[39] J.P. Kotter and J.L. Heskett, Corporate culture and performance, The Free Press, New York, NY, 1992.

[40] J. Landis and G.G. Koch, The measurement of observer agreement for categorical data, *Biometrics* 33 (1977) 159-174.

[41] H.Y. Lee, H.W. Jung, C.S. Chung; J. M. Lee, K. W. Lee and H. J. Jeong, Analysis of inter-rater agreement in ISO/IEC 15504-based software process assessment, in: Proceedings of the 2nd Asia-Pacific Conference on Quality Software, 2001, pp. 341-348.

[42] M.A. Lyles, An analysis of discrimination skills as a process of organizational learning, The Learning Organization 1 (1) (1994) 23-32.

[43] R.R.. Macala, L.D. Jr. Stuckey and D.C. Gross, Managing domain-specific, product-line development, IEEE Software 13 (3) (1996) 57-67.

[44] M. Mannion, Organizing for software product line engineering, in: Proceedings of the 10th International Workshop on Software Technology and Engineering Practice, 2002, pp. 55 –61.

[45] M. Marquardt and A. Reynolds, The global learning organization, Irwin, Illinois, 1994.

[46] J.E. Mathieu and D. Zajac, A review and meta-analysis of the antecedents, correlates, and consequences of organizational commitment, Psychological Bulletin 108 (1990) 171-94.

[47] J.D. McGregor, Software product lines, Journal of Object Technology 3 (3) (2004) 65-74.

[48] F. J. Medina, L. Munduate, M.A. Dorado and I. Martínez, Types of intra-group conflict and affective reactions, Journal of Managerial Psychology 20 (3/4) (2005) 219-230.

[49] J.C. Nunnally, and I.A. Bernste, Psychometric theory, Third ed.: McGraw Hill, New York, 1994.

[50] H. Nyqusit, Sensitivity analysis of empirical studies, Journal of Official Statistics 8 (2) (1992) 167-182.

[51] C. O'Reilly, and J. Chatman, Culture as social control: corporation, cults, and commitment, Research in Organizational Behavior 8 (1996) 157-200.

[52] A. Osterhof, Classroom applications of educational measurement, Prentice Hall, NJ. 2001.

[53] B. Paterson, Still plausible stories: a review of Alfred Chandler's classics, Academy of Management Review 13(4) (1988) 653-656.

[54] R. Rosen, Strategic management: an introduction. Pitman, London, UK, 1995.

[55] A. Saltelli, K. Chan and M. Scott, Sensitivity analysis, probability and statistics series. John Wiley & Sons, NY, 2000.

[56] E. H. Schein, Organizational psychology, Prentice Hall, 1988.

[57] C. H. Schwepker, Ethical climate's relationship to job satisfaction, organizational commitment, and turnover intention in the sales force, Journal of Business Research 54 (2001) 39-32.

[58] W. R. Scott, Institutions and organizations, Sage Publications, CA, 1995.

[59] I.M. Sobol, Sensitivity estimates for nonlinear mathematical models, Mathematical Modeling Computers 1 (4) (1993) 407-414.

[60] J. Stevens, Applied multivariate statistics for the social sciences, Hillsdale, NJ, 1986.

[61] D.W. Straub, Validating instruments in MIS research, MIS Quarterly 13 (2) (1989) 147-169.

[62] A. Todd, Managing radical change, Long Range Planning 32 (2) (1999) 237-44.

[63] P. Toft, D. Coleman and J. Ohta, A cooperative model for cross-divisional product development for a software product line, in: Proceedings of the 1st International Conference on Software Product Lines, 2000, pp. 111-132.

[64] F. *van der* Linden, Software product families in Europe: The Esaps & Café projects, IEEE Software 19 (4) (2002) 41-49.

[65] F. *van der* Linden, J. Bosch, E. Kamsties, K. Känsälä and H. Obbink, Software Product Family Evaluation, in: Proceedings of the 3rd International Conference on Software Product Lines, 2004, pp. 110-129.

[66] A.H. *van de* Ven and D.L. Ferry, Measuring and assessing organizations, John Wiley & Son: NY, 1980.

[67] M. Verlage and T. Kiesgen, Five years of product line engineering in a small company, in: Proceedings of the 27th International Conference on Software Engineering, 2005, pp. 534 – 543.

[68] A. *von* Eye, E.Y. Mun, Analyzing Rater Agreement Manifest Variable Methods, LEA Publishers, London, 2005.

[69] J.A. Walls and R.R. Callister, Conflict and its management, Journal of Management 21 (3) (1995) 515-558.

[70] D.M. Weiss and C.T.R. Lai, Software product line engineering: a family based software development process, Addison Wesley, 1999.

[71] D.C. Wilson and R.H. Rosenfeld, Managing Organizations, McGraw-Hill, 1990.

[72] A.M. Wilson, Understanding organizational culture and the implication for corporate marketing, European Journal of Marketing 35 (3/4) (2001) 353-67.

# Appendix-I

## Organizational Factors (Measuring Instrument)

### Organizational Structure

1. Roles and responsibilities of individual and groups are well defined and documented in the organization.
2. Organizational structure supports the software product line.
3. A strong and open communication channel among various entities of the organization is present.
4. The organization's strategic plans define how it will achieve the technological capability to successfully adopt the concept of software product lines.

### Change Management

5. The organization has a well-defined change management plan to switch from single product development to a product line.

6. The change management plan is well communicated to all employees of the organization.
7. The changes in the organization to institutionalize software product line in the organization are well accepted by the employees.
8. The resistance to change in the organization is gradually decreasing.

**Organizational Culture**

9. The workforce understands, and is committed to the vision, values and goals of the organization.
10. The organizational culture supports reusability of software assets.
11. Management encourages new ideas and improvement plans.
12. It is not difficult for a new employee to settle down in the working environment of the organization.

**Organizational Commitment**

13. Employees feel a sense of ownership for this organization rather than being just an employee.
14. I would accept other job assignment in order to keep working for this organization.
15. Over the past three years the organization as a whole entity is steadily moving towards software product line approach in order to achieve strategic objectives.
16. Employees consider software product line as a vital entity to achieve its long-term goals.

**Organizational Learning**

17. Formal and informal mechanisms are used to disseminate learning and knowledge within organization.
18. Necessary training is provided to employees about software product line concept.
19. Organization learns from its experience and lessons and avoids making mistakes again and gain.
20. Research and development in software product line is a continuous process in the organization.

**Conflict Management**

21. The organization has setup a policy to handle conflicts within the organization.
22. Management supports positive and constructive conflict.
23. Personal conflicts are major hurdle in the progress of the organization.
24. Employees handle their conflict at their own.

**Software Product Line Performance**

1. Over the past three years, the organization is able to reduce cost, product defects and development time of software products.
2. The sales of the organization have steadily increased over the past three years and the organization is able to attract new customers and launch new products.
3. Software product line is playing a significant role in achieving the business goals of the organization.
4. Financial analysis shows a progressive growth over the last three years due to software product lines.